\documentclass[11pt,twoside]{article}

\usepackage{asp2014}

\aspSuppressVolSlug
\resetcounters

\begin{document}

\title{IAU CPS Tools to Address Satellite Constellation Interference}

\author{Michelle Dadighat,$^{1,5}$ Meredith L. Rawls,$^{2,5}$ Siegfried Eggl,$^{3,5}$ Mike Peel,$^{4,5}$ and Constance E. Walker$^{1,5}$}
\affil{$^1$NSF's NOIRLab, Tucson, Arizona, USA; \email{michelle.dadighat@noirlab.edu}}
\affil{$^2$University of Washington, Seattle, WA, USA}
\affil{$^3$University of Illinois at Urbana-Champaign, Champaign, IL, USA}
\affil{$^4$Imperial College London, Blackett Lab, Prince Consort Road, London, UK}
\affil{$^5$IAU Centre for the Protection of the Dark and Quiet Sky from Satellite Constellation Interference (CPS)}

\paperauthor{Michelle Dadighat}{michelle.dadighat@noirlab.edu}{}{NSF's NOIRLab}{}{Tucson}{AZ}{}{USA}
\paperauthor{Meredith L. Rawls}{mrawls@uw.edu}{0000-0003-1305-7308}{University of Washington}{Department of Astronomy/DiRAC/Vera C. Rubin Observatory}{Seattle}{WA}{}{USA}
\paperauthor{Siegfried Eggl}{eggl@illinois.edu}{0000-0002-1398-6302}{University of Illinois at Urbana-Champaign}{Department of Aerospace Engineering/Department of Astronomy}{Champaign}{IL}{}{USA}
\paperauthor{Mike Peel}{m.peel@imperial.ac.uk}{0000-0003-3412-2586}{Imperial College London}{Blackett Lab}{London}{}{SW7 2AZ}{UK}
\paperauthor{Constance E. Walker}{connie.walker@noirlab.edu}{0000-0003-0064-4298}{NSF's NOIRLab}{IAU CPS}{Tucson}{AZ}{}{US}



\begin{abstract}
The IAU Centre for the Protection of the Dark and Quiet Sky from Satellite Constellation Interference (CPS), established in early 2022 and co-hosted by NSF's NOIRLab and the SKA Observatory, was created to unify efforts to work towards mitigating some of the effects of satellite constellations on astronomy. SatHub, one of the four sub-groups of CPS, focuses on software and related tools to aid observers and industry partners in addressing some of the issues caused by commercial satellite constellations.
\end{abstract}



\section{Introduction}
SatHub, part of the IAU CPS, is a community-driven group that is working on, among other goals, developing and promoting different software tools to help address the issue of satellite constellation interference as it affects astronomical observations (\citet{rawls21-satcon2,peel23-iaus385}). Currently, SatHub software efforts are concentrated in two main areas: SatChecker\footnote{SatChecker API Documentation: \url{https://satchecker.readthedocs.io/en/latest/}.}, a satellite ephemeris prediction service which is nearing a beta phase, and a satellite brightness observation database, still in early development.

\section{SatChecker}
One of the primary tools needed to mitigate some of the effects of commercial satellite constellations is a robust ephemeris service that can provide accurate satellite positions as well as estimated brightness (calculated using Lumos-Sat\footnote{Lumos-Sat: \url{https://github.com/Forrest-Fankhauser/lumos-sat}}, see \citet{fankhauser23} ). This can reduce or remove the need to recreate satellite interference predictions across multiple observatory locations. The SatChecker service will be available via API and web interfaces with the ability to provide satellite passage information for specific pointing/FOV/exposure times, as well as general ephemeris information for specific satellites. 

Currently SatChecker is able to provide satellite position information, on-sky velocity, and whether the satellite is illuminated when searching by satellite name or NORAD catalog ID over a requested date range. The FOV work is pending future funding, but will include the ability to find which satellites are crossing the user specified field of view centered around a given RA/Dec at a specified time window.

Another way that SatChecker can assist with mitigation efforts is make it easier to observe the satellites themselves as part of specific observing campaigns. For example, Starlink Gen2 Mini and Amazon Kuiper satellites have observing campaigns planned, and SatChecker could be used to easily and accurately predict where to find specific satellites and whether they are suitable for observing at a particular time.

\articlefigure[width=.75\textwidth]{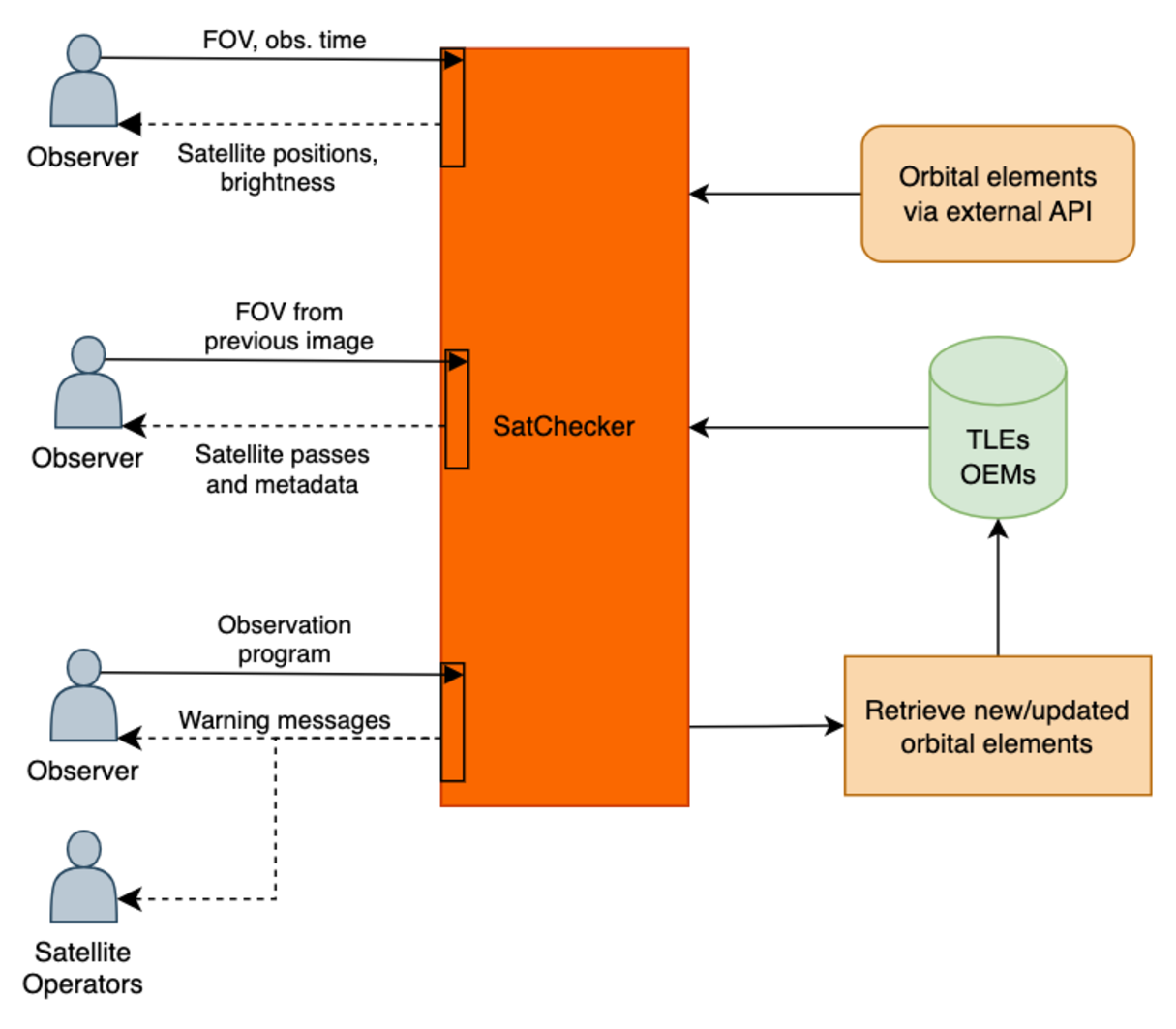}{fig1}{SatChecker}

SatChecker will also utilize multiple data sources providing orbital data in both two-line element (TLE) and orbital ephemeris message (OEM) formats to both predict future passes and provide archival information. Using SatChecker for archival searches will utilize the orbital data closest to the requested time, since satellite positions are useful for up to two weeks on either side of the TLE epoch date, although precision degrades much faster than that. Additionally, SatChecker could also be used to conduct simulations of how satellite constellations can impact specific science cases, such as Rubin Observatory’s LSST and associated follow-up observations.

\section{Satellite Observation Repository}
Another focus of SatHub’s development efforts is a data repository to collect satellite brightness observations - visual, optical, radio, and the images themselves. Currently work has begun on a website to handle image-only uploads (Trailblazer\footnote{Trailblazer: \url{https://trailblazer.dirac.dev/}.}), but we will also be working on developing a repository to collect all brightness observation measurements and make them easily accessible for review or download. 

\articlefigure[width=.75\textwidth]{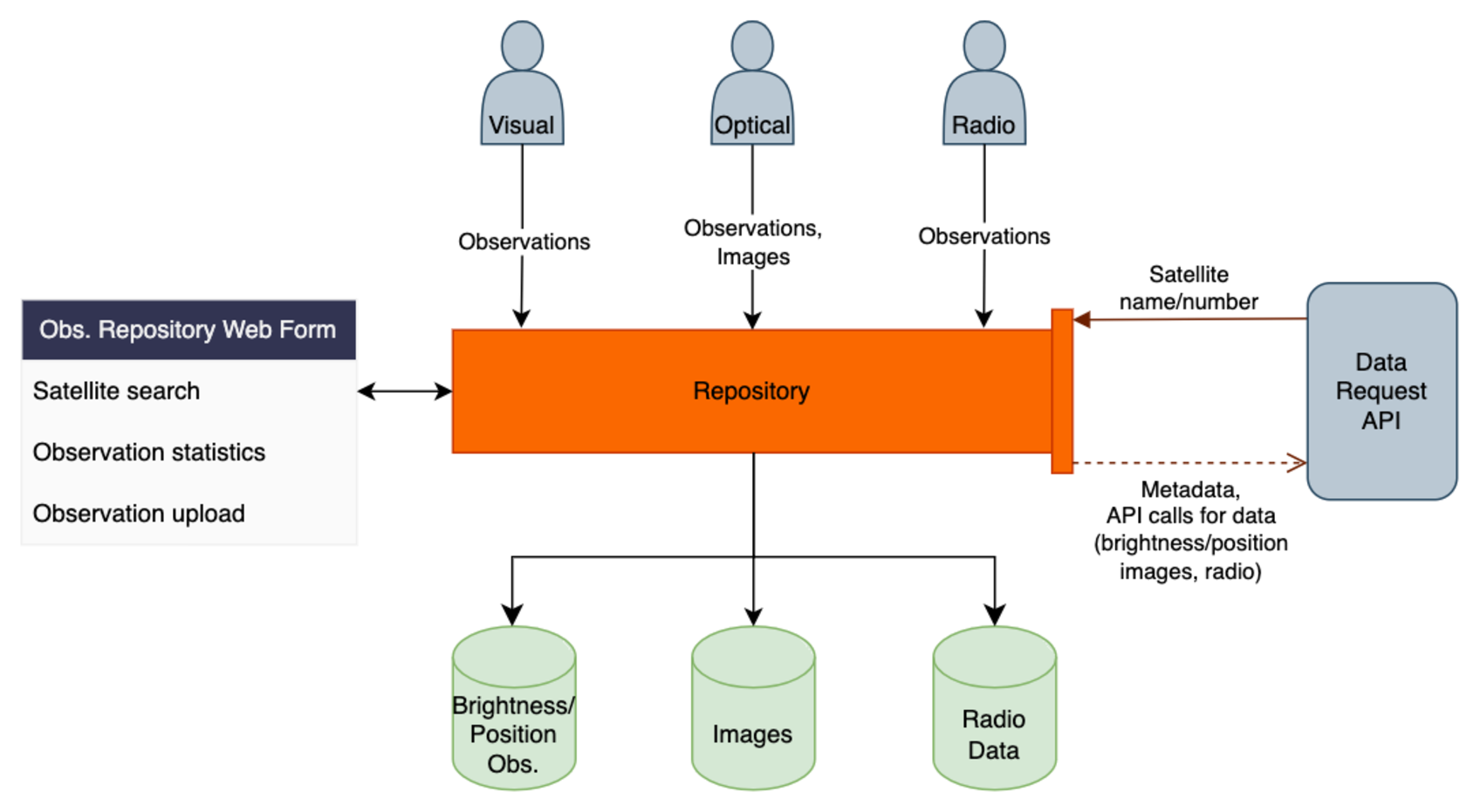}{fig2}{Satellite Observation Repository}

The satellite observation repository will provide an API and web form for the upload of optical and visual brightness and position measurements using a standardized data format. Measurements can be accompanied by corresponding image data, but this is not required. It will also support the collection of radio (and eventually spectral) data of satellite passes.

This data can be used to simulate satellite constellation interference on different science programs, to characterize satellite constellations over time, or to verify whether brightness mitigation efforts by satellite operators are effective. Eventually, this repository could also be integrated with SatChecker to gather brightness measurements from images uploaded to Trailblazer.

\bibliography{P925.bib}

\end{document}